\documentclass[12pt]{article}
\begin{document}
\begin{center}
{\LARGE Lepton Masses and Mixings in Next-to-minimal \\
Supersymmetric SO(10) GUT}
\\ \mbox{} \\
{M. MALINSK\'Y}\footnote{E-mail: malinsky@sissa.it}
\\ \mbox{} \\
Scuola Internazionale Superiore di Studi Avanzati (S.I.S.S.A.)\\
Via Beirut 2-4,  
34014 Trieste, Italy\\ 
\end{center}

\abstract{
A simple extension of the minimal renormalizable supersymmetric SO(10) grand
unified theory by adding a 120-dimensional Higgs representation is examined. This brings new
antisymmetric contributions to the relevant quark and lepton mass sum rules and leads to
a better fit of the measured values of lepton masses and mixings together with a natural completion of the
renormalizable Higgs sector within the SUSY SO(10) framework. 
}

\section{Introduction}
The class of supersymmetric grand unified theories (GUT) based on the $SO(10)$ gauge group 
seems to be one of the most promissing frameworks to describe the physics beyond the Standard
model, including massive neutrinos. Though the scale at which the GUT symmetry should be realized is very
large ($\sim 10^{16}$~GeV)
there could be observable consequences of such scenarios at the laboratory energies, be it the
tiny neutrino masses, measurable proton decay rate, anomalous electric dipole moments and other
phenomena.  The high scale symmetry propagates into the low-energy observables by means of
effective relations among quantities which are in general uncorrelated within the SM framework.  
This makes such scenarios quite predictive and thus very attractive from the point of view of the
low-energy phenomenology. 
In this talk we give a short overview of the minimal renormalizable SUSY
SO(10)\cite{Clark:ai,Aulakh:1982sw,Aulakh:2003kg}, in
particular the effective mass sum rules for quark and lepton mass matrices arising from the GUT-scale
physics
and their consequences on the leptonic sector, namely the predictions for the neutrino
masses and the PMNS lepton mixing matrix. Then we define a simple renormalizable extension 
of the minimal model by including one additional (almost decoupled) 120-dimensional Higgs
multiplet.
We argue that even a tiny admixture of its bidoublet components within the light MSSM Higgs doublets can
lead to substantial effects in the predicted values of the neutrino masses and PMNS mixing angles. 
 
\section{Minimal renormalizable SUSY SO(10) model}
In the past few years one may notice a 'renaissance'
\cite{Bajc:2002iw,Aulakh:2003kg,Bajc:2004xe,Bajc:2004fj,Mohapatra12,Bertolini:2004eq,Dutta:2004hp} of the
renormalizable SUSY SO(10) model. It was shown that this framework can accommodate good  R-parity
conserving $SO(10)\to SM$ breaking patterns being more
constrained than any realistic GUT model based on the $SU(5)$ gauge group\cite{Aulakh:2003kg}. Moreover, there is
an intriguing relationship \cite{Brahmachari:1997cq} between the (approximate) maximality of the
atmospheric mixing
in the leptonic sector and the $b$-$\tau$ unification provided the neutrino mass matrix is dominated by
the type-II seesaw contribution. \\ \\
{\bf Structure of the minimal renormalizable SUSY SO(10)}\\
One of the most appealing features of any $SO(10)$ GUT model is the fact that the SM fermions of
each generation (including the right-handed neutrinos) reside in one irreducible 16-dimensional representation of
the gauge group, the spinorial $16_{F}$. Concerning the Yukawa part of the
superpotential, the matter bilinear $16_{F}\times 16_{F}$ can couple at the renormalizable level
only to three types of Higgs multiplets, namely the 10-dimensional vector multiplet $10_{H}$, the
126-dimensional antiselfdual 5-index antisymmetric tensor $\overline{126}_{H}$ and the 120-dimensional three-index
antisymmetric tensor $120_{H}$. It was shown\cite{Aulakh:2003kg,Mohapatra12} that in order to
obtain a
realistic  leptonic spectrum it is sufficient to consider only the  $10_{H}$ and
$\overline{126}_{H}$ Higgs multiplets. (In addition the 210-dimensional
four-index antisymmetric tensor $210_{H}$ is needed to break properly the $SO(10)$ group down to
the MSSM and mix the $10_{H}$ and $\overline{126}_{H}$ to generate
the left-handed triplet VEV entering the seesaw formula together with the proper mixings among  the
components entering the two light Higgs doublets of the MSSM).  
\\
\\
{\bf Sum rules for quark and lepton mass matrices in MRM}\\
Inspecting the $SU(3)_{c}\times SU(2)_{L}\times U(1)_{Y}$ structure of these multiplets one sees
that the quark and lepton mass matrices obey
\begin{eqnarray}\label{1}
M_{u}& =& Y_{10}v^{10}_{u}+Y_{126}v^{126}_{u}\nonumber \qquad
M_{d} = Y_{10}v^{10}_{d}+Y_{126}v^{126}_{d}\nonumber \\
M_{l}& =& Y_{10}v^{10}_{d}-3 Y_{126}v^{126}_{d}\quad\,
M^{D}_{\nu} = Y_{10}v^{10}_{u}-3 Y_{126}v^{126}_{u} \\
M^{R}_{\nu}& \propto & Y_{126}\langle (1,1,0)_{\overline{126}}\rangle \nonumber \qquad\,\,
M^{L}_{\nu} \propto Y_{126} \langle (1,3,+2)_{\overline{126}}\rangle \nonumber 
\end{eqnarray}
provided $Y_{10}$ and $Y_{126}$ are the (symmetric) Yukawa matrices parametrizing the coupling of the
matter bilinear to the relevant Higgs multiplets and
$v^{10,126}_{u,d}$ are the VEVs of bidoublets contained in $10_{H}$ and
$\overline{126}_{H}$.
\\
\\
{\bf Lepton mixing in MRM with dominant type-II seesaw}\\
Assuming that the type-II contribution dominates the neutrino mass-formula one can obtain the
following sum-rules for the charged lepton and neutrino mass matrices\cite{Mohapatra12} (in the basis
where $M_{d}$
is diagonal $\equiv D_{d}$; the tilde denotes the mass matrix normalized to its largest eigenvalue)
\begin{equation}\label{2}
k \tilde{M}_{l} = V_{CKM}^{T} \tilde{D}_{u}V_{CKM}+r \tilde{D}_{d}\qquad
M_{\nu}\propto  \tilde{M}_{l}-\frac{m_{b}}{m_{\tau}}\tilde{D}_{d}
\end{equation}
Here $k$ and $r$ are functions of the $v$-parameters in (\ref{1}). Looking at (\ref{2}) one can
appreciate the predictivity of the model: $i)$ It is very nontrivial to get a good fit of the charged
lepton mass ratios at the LHS of (\ref{2}) by varying only the quark masses
and mixings within their experimental ranges and using the freedom in $r$ and the remaining 6 complex
phases (in $\tilde{D}_{x}$) on the RHS. $ii)$ Whenever one finds a region in the
parametric space where the charged lepton mass ratios fit well, the neutrino mass matrix
is known up to just one
phase. Thus the model is very predictive in the neutrino sector. Moreover, if $m_{b}$ approaches
$m_{\tau}$ (and the relative phase of the two terms is adjusted properly),  the 33-entry of $m_{\nu}$
is comparable with the other entries in the 23 sector, what leads to the almost
bimaximal structure of the $U_{PMNS}$\cite{Brahmachari:1997cq}.  
Coming to the numerical analyses\cite{Mohapatra12,Bertolini:2004eq} (with the CP-phases switched
off for simplicity; their
effects were shown to be subleading in most cases\cite{Mohapatra12} usually worsening the fit of the
charged lepton formula), the following predictions are obtained at the 1-$\sigma$
level \cite{Mohapatra12,Bertolini:2004eq}: $|U_{e3}|\geq 0.15$, $\sin^{2}2\theta_{13}\geq 0.85$,
$\sin^{2}2\theta_{23}\leq 0.97$. The lower bound for $|U_{e3}|$ turns out to be very rigid and can be
a 'smoking gun' of the model. Moreover, the solar mixing tends to be too large, while the atmospheric is
never maximal.
 
\section{Extending the minimal renormalizable model}
Though the minimal model predictions are in a reasonably good agreement with the experimental data by
extending the analysis to 2-$\sigma$ level one
may ask whether some extensions of MRM may perform better. However, often the price to be
paid is the lack of predictivity and one should look for extensions that are
constrained enough to remain as predictive as possible.  
\\
\\
{\bf Adding a 120-dimensional Higgs representation}
\\
One of (few) such renormalizable generalizations of the MRM is the scenario with one additional
quasidecoupled 120-dimensional
 Higgs representation (to which we refer as the next-to-minimal renormalizable model,
NMRM)\cite{Bertolini:2004eq}.
The key observation is that the $120_{H}$ multiplet can be naturally heavier than the GUT scale, because
it does not participate at the GUT-symmetry breaking. Its scalar mass
parameter $M_{120}$ is not constrained by potential-flattness conditions  and
can be 
naturally as large as the cut-off, be it the Planck scale. This means that the weights  of the
bidoublet components entering the weak-scale MSSM Higgs
doublets may be naturally suppressed with respect to those coming from $10_{H}$ and
$\overline{126}_{H}$.
Therefore the relations (\ref{1}) are only slightly modified thus preserving most of the good features
of the MRM. Let us 
write down the new quark and lepton mass formulae :
\begin{eqnarray}\label{5}
M_{u}& =& Y_{10}v^{10}_{u}+Y_{126}v^{126}_{u}+Y_{120}v^{120}_{u}\nonumber \qquad
M_{d} = Y_{10}v^{10}_{d}+Y_{126}v^{126}_{d}+Y_{120}v^{120}_{d}\nonumber \\
M_{l} &=& Y_{10}v^{10}_{d}-3 Y_{126}v^{126}_{d}+Y_{120}v^{120}_{l} \quad \,\,
M_{\nu} \propto  Y_{126}\langle (1,3,+2)_{\overline{126}}\rangle \label{2.5}
\end{eqnarray}
The inequality $M_{120}\gg M_{GUT}$  translates into $v^{120}_{x}\ll
v_{u,d}^{10,126}$. Equivalently, one can write (diagonalizing the quark mass matrices by means of
biunitary transformations $M_x=V^R_x
D_x {V^L_x}^T,~x=u,d$ and denoting $W\equiv{V^R_u}^T V_d^R$ , 
$V_{CKM}\equiv {V_u^L}^T V_d^L$ and 
$Y_{120}'\equiv{V^R_d}^T Y_{120}V^L_d$)
\begin{eqnarray}
\label{6}
k {V^R_d}^T\tilde{M_l}V^L_d &= &W^T\tilde{D}_u V_{CKM} + r \tilde{D}_d + Y_{120}'
(k \varepsilon_l-\varepsilon_u-r \varepsilon_d) \nonumber \\
\label{7}
M_\nu & \propto & \tilde{M_l} -\frac{m_{b}}{m_{\tau}}\tilde{M_d}+Y_{120}
\left(\frac{m_{b}}{m_{\tau}}\varepsilon_d-
\varepsilon_l\right)
\end{eqnarray}
with $\varepsilon_{u,d,l}\equiv v_{u,d,l}/m_{t,b,\tau}$. Since $Y_{120}$ is antisymmetric, the
$M_{u,d,l}$ are no longer symmetric and the unknown right-handed quark mixing matrix $W$ appears
at the RHS of (\ref{6}). However, since this setup is a perturbation of the MRM one can expand the $W$ matrix
around $V_{CKM}$ by means of the small parameters in the game (neglecting the CP phases):
$W=V_{CKM}+2\left(-\varepsilon_u Z_u V_{CKM}+ \varepsilon_d V_{CKM} Z_d\right) +
{O}(\varepsilon_x^{2}) $ where the $Z_{x}$ matrices are given by
$(Z_{x})_{ij}={(Y_x')_{ij}}/\left[{(\tilde{D}_x)_{ii}+(\tilde{D}_x)_{jj}}\right]$ and 
$Y'_{u} \equiv  V_{CKM} Y_{120}' V_{CKM}^T$, $Y'_{d} \equiv Y_{120}'$. 
Therefore, for small $\varepsilon_{x}$ the predictions of this model are expected to be close to those of
MRM. It can be shown that the tiny antisymmetric corrections change  the
PMNS angles linearly in $\varepsilon's$ while the masses are affected at the second order,
what makes the perturbative method self-consistent and the fit of
the charged lepton mass matrix stable enough to preserve the good features of the MRM.
On the other hand, the 1-2 entries of the $Z_{x}$ matrices can be strongly enhanced by the small 
'$\tilde{D}_{11,22}$' terms in the denominator.
\\
\\
{\bf Lepton mixing in NMRM with dominant type-II seesaw}
\\
The numerics shows\cite{Bertolini:2004eq} that even for $\varepsilon \sim 10^{-3}$ the non-decoupling
effects 
of the $120_{H}$ contributions to the $PMNS$ angles can reach several tens of percent. For instance
the MRM lower bound for $|U_{e3}|$ can be relaxed to about $|U_{e3}|> 0.1$, while the
atmospheric angle can be maximal and the solar bound is changed to about
$\sin^{2}2\theta_{13}>0.75$, all this at 1-$\sigma$ level even with the CP-phases switched off. 
\section{Conclusions}
We have argued that the lepton sector predictions of the minimal renormalizable SUSY SO(10) model
are very sensitive to the magnitude of the antisymmetric Yukawa structure be it an additional small
renormalizable coupling of $120_{H}$ Higgs multiplet to the matter bilinear (or an effective vertex
generated by dynamics beyond the GUT-scale). Therefore such terms should be taken into serious
consideration
when discussing the phenomenological implications of such class of grand unified theories. 
\section*{Acknowledgements}
I am very grateful to Stefano Bertolini for the discussions and encouragement throughout the preparation
of this contribution and to Michele Frigerio for useful comments. It is also a pleasure to thank the
organizers of the Particles, Strings and Cosmology 2004 (PASCOS'04) conference for warm hospitality at the
Northeastern University in Boston.
 
\end{document}